 \documentclass[final,5p,twocolumn,authoryear]{elsarticle}
\usepackage{amssymb}
\usepackage{amsmath}
\usepackage{bm}
\usepackage{braket}
\usepackage{lineno}
\usepackage{graphicx}\graphicspath{{figures}}
\usepackage{subcaption}
\usepackage{hyperref}
\usepackage{lettrine}
\journal{  }

\begin{document}

\begin{frontmatter}

\title{Diffusion and turbulence in phase-space and formation of phase-space vortices}
\author[first]{Allen Lobo}\ead{allen.e.lobo@outlook.com}
\affiliation[first]{organization={Sikkim Manipal Institute of Technology},
            addressline={Sikkim Manipal University}, 
            city={Majitar},
            postcode={737136}, 
            state={Sikkim},
            country={India}.}
\author[first]{Vinod Kumar Sayal}

\begin{abstract}
In this work, the recently introduced fluid-like treatment of the phase-space has been further extended and some interesting outcomes have been presented. A modified form of the Vlasov equation has been presented which describes the diffusion of the phase-space density. This anisotropic diffusion is analysed and the flow of the phase-space probability field is shown. Growth of phase-space vortices is then shown due to increased turbulent-like flow, which is marked by the dominating inertial flow above the diffusive flow. The nature of this flow is judged by using a parameter for the phase-space. It is then shown that the formation of phase-space vortices is due to growth of turbulent-like flow in the phase-space. On the bases of the diffusion parameters, the vorticity field transport of the hydrodynamic phase-space is studied and a Schamel-KdV form of the vorticity transport equation is derived, suggesting solitary modes of the phase-space vorticity waves.
\end{abstract}


\end{frontmatter}


\section{Introduction}
\lettrine[]{M}{}any particle systems, such as plasmas, are well-known to exhibit fluid-like behavior, which treats such systems as a continuum and depicts various fluid-analogous characteristics in order to explain the collective dynamics of the system. One such characteristic behavior, which is majorly responsible for formation of coherent structures (including vortices in fluids) is the presence of turbulence. Turbulence by itself is not an analytically well-defined nature of fluid flows\citep{Eames2011NewFlows}, but is principally recognizable by at-least two distinct phenomena associated with it\citep{batchelor1967introduction, Majumdar2003AnGeneration, Tryggeson2007AnalyticalEquation} --
\begin{itemize}
    \item dominance of the inertial flow of the fluid over the viscous/diffusive flow, and
    \item growth of strong vorticity fields followed by formation of vortices in these regions.
\end{itemize}

The presence of fluid turbulence is usually indicated by dimensionless parameters like the Reynolds number \citep{Reynolds1883XXIX.Channels}, which portrait the  dominance of inertial flow over the viscous slip of the fluid layers on top of each other. However, the minimum criteria for recognizing a turbulent flow is the sufficient dominance of the inertial flow over the diffusive/viscous flow of the fluid in the local region. Turbulence typically occurs for large values of the Reynolds number. Formation of vortical structures are well-known to result from this turbulent state of the fluid \citep{Majumdar2003AnGeneration, Dyachenko2008VortexWalls, Shi2023VortexNumber}.

The kinetic approach to the study of systems such as collisionless plasmas, on the other hand, explore the dynamical evolution of the phase-space of the species. This evolution is described by the well-known Vlasov equation \citep{Vlasov1938}, along-with force-field equations which collectively depict the various interaction modes occurring within the system. The kinetic theory studies the temporal evolution of the phase-space probability density and predicts the nature and behavior of the system characteristics, such as the particle density and associated force fields. It is well-known that the kinetic theory presents a more holistic description of the system behavior, and is therefore often employed for the analyses of some complex physical phenomena which the hydrodynamical model fails to predict. One such phenomena is the particle-trapping solitary wave-modes, known as Bernstein-Greene-Kruskal (BGK) modes \citep{Bernstein1957}. These are also known as phase-space holes or simply as phase-space vortices. These waves exhibit vortical motion of the trapped particle in the phase-space, which can be seen as closed trajectories \citep{hutch2017, Lobo2024TheoryPlasmas, Lobo2024StrongHoles}.

The particle phase-space bears resemblance with a two-dimensional fluid surface, due to the identical nature of the Vlasov equation and fluid continuity equation \citep{Berk1970PhaseObservations, schamel1986electron}. In a recent work by \citet{Lobo2024TheoryPlasmas}, this analogy between the particle phase-space and ordinary two-dimensional fluids has been significantly developed. Specifically, using the phase-space hydrodynamic model developed in their work, \citet{Lobo2024TheoryPlasmas} show that the phase-space of many-particle systems, in particular the one-dimensional electrostatic plasma, can be studied in close analogy to a two-dimensional ordinary fluid, and that this analogy exists beyond the Vlasov-Poisson equation system. Using this fluid-like modeling, phase-space holes -- which are nonlinear kinetic structures found in the phase-space of collision-free plasmas appearing as Debye-scale cusp-like depressions of probability densities, were shown to exhibit vortex-like nature in the phase-space. These phase-space holes are well-known to form through two-stream-like instabilities in the plasma, and are nonlinear solutions (Bernstein-Greene-Kruskal (B.G.K.) modes) of the Vlasov-Poisson equation system \citep[see][and references therein]{Bernstein1957, schamel1986electron, schamel2012cnoidal, hutch2017}.

\indent The problem of phase-space holes has been addressed by various authors following various approaches chiefly the B.G.K. differential approach \citep{Schamel1979, schamel1982kinetic, schamel1986electron, schamel2012cnoidal,jenab2021ultrafast}, the integral approach \citep{Turikov1984, Chen2005, Aravindakshan2021StructuralPlasma, lewis1984, Abraham-Shrauner1998}, the entropy-based approach as followed by \citet{Dupree1982}, and very recently the phase-space hydrodynamic model \citep{Lobo2024TheoryPlasmas}. These works, along-with numerous simulation and experimental studies \citep{morse1969one, Saeki1979, Guio2003, Lynov1985Phase-SpaceHoles, Sayal1994ThePlasma, Eliasson2006, Lotekar2020MultisatelliteInstability, Wang2022MultisatelliteSheet}, have collectively described various aspects of phase-space hole behavior, including the origin of holes from various trapping scenarios, their structure in phase-space, solitary wave solution and a nonlinear dispersion relation. However, the focus on the vortical nature of phase-space holes, which is presented by the phase-space hydrodynamic approach, describes interesting analogies between hole characteristics and fluid vortices. This approach also presents a well-derivable set of the hole phase-space structure, both in the trapped and free regions, which were initially required to be assumed \citet{schamel1986electron}. 

Inspired by the fluid-analogous phase-space hydrodynamic model, in this paper we aim to address an interesting question -- analogous to fluid vortices, whether the phase-space holes develop in the plasma from turbulent-like flows, and if so, what mathematical analyses describes this kinetic phenomena. This problem serves as the chief motivation for this work. In this article, the phase-space hydrodynamic model is further developed to explore the development of phase-space vortices. For the same, the fluid-like dynamics of the phase-space probability distribution function is observed. This flow-like nature of the phase-space probability is mathematically analyzed and a diffusive-flow is reported. An idea of statistical turbulence is introduced in analogy to conventional fluid physics and formation of phase-space holes is observed using numerical simulations. The analytical results are then employed to study the phase-space vortex transport phenomena and the vorticity transport equation is derived for the phase-space hydrodynamic model. This equation is found to be somewhat identical to the well-known Schamel-Korteweg-De Vries equation \citep{Schamel1973AElectrons}, which describes various solitary wave-modes in the system.

This paper is organized as follows: in section 2 the phase-space hydrodynamic model is discussed and the theory is extended in order to present diffusive flow of the phase-space density fluid. Turbulence is then analytically measured as a difference in the magnitudes of the inertial and diffusive flow rate. This is followed by section 3 in which 
the phase-space of one-dimensional electrostatic plasmas is numerically simulated using some well-known techniques \citep{Cheng1976, turikov1978computer, Lobo2024AccelerationPlasma} and the mathematical analyses of section 2 are established. Section 4 then presents the vorticity transport equation using the diffusion parameters presented in section 2, and develops a dispersion form of the vorticity transport equation which suggests solitary wave-like propagation. Section 5 then presents the concluding remarks. The findings presented in this paper, apart from exhibiting vortex formation in phase-space due to existing 'turbulent' flow, also describe a kinetic diffusion of probability due to existing dynamical fields in the system, and hence present some new findings in the domain of many-particle kinetics.

\section{Diffusion and turbulence in the phase-space hydrodynamical analogy}\label{sec2}
\subsection{The phase-space hydrodynamic model}
\indent Fundamentally, the phase-space dynamics of a classical collision-free system is governed by the Vlasov equation. The force term $F(x)$ can be appropriately replaced by system-specific governing equations of the present force fields. In the case of an electrostatic plasma, the well-known Poisson equation can be inserted into the Vlasov equation, constructing the Vlasov-Poisson equation,
\begin{equation}\label{Vlasov_equation}
    \frac{\partial f}{\partial t} + v_x\frac{\partial f}{\partial x} - \frac{q}{m}\frac{\partial \phi}{\partial x}\frac{\partial f}{\partial v_x}=0,
\end{equation}
 where the phase-space co-ordinates $(x,v_x)$ represent the position and velocity of a phase-space element. Here, $\phi(x)$ represents the existing spatial scalar potential field in the system consisting of particles of mass $m$, such that the electrostatic field, $E(x) = -\partial_x\phi(x)$, and $q$ is the particle charge. However, the phase-space, in its fluid analogue, can also be represented using its streaming function $\psi(x,v_x)$, which, according to \citet{Lobo2024TheoryPlasmas} is a modified Hamiltonian of the system,
\begin{equation}\label{Stream_function}
    \psi(x,v_x) = \frac{\tau}{m}\left(\frac{1}{2}mv_x^2 + q\phi(x)\right).
\end{equation}
In the above equations, $\tau$ represents a characteristic response time of the particle in the classical system, and is a constant for the system. The streaming function conveniently describes the fluid-analogous nature of the phase-space and incorporates the Vlasov-Poisson equation as the continuity equation. It also defines phase-space vortices by a simple relation --
\begin{equation}\label{streamfuncitonconditionforhole}
    \psi(x,v_x)\quad\begin{cases}
        \quad<0\quad&\text{rotational flow (vortex),}\\
        \quad\geq 0 \quad & \text{non-rotational flow.}
    \end{cases}
\end{equation}
As is clear, the above equation (\ref{streamfuncitonconditionforhole}) actually associates the vortical flow with particle trapping. The phase-space plane $(x,\tau v_x)$ depicts the dynamics of the probability distribution of particles of the system, along position and velocity. Under the influence of conservative force fields, these particles strictly follow phase-space trajectories of constant Hamiltonian, such that
\begin{equation}
     mv_x dv_x - qE(x)dx = 0.
\end{equation}
However, in the fluid phase-space model, the fluid nature of the phase-space elements is analysed in terms of position-space $(\hat{x})$ and ($\tau$ times) the velocity space $(\hat{v}_x)$. Here, the velocity axis $\tau v_x$ refers to the extent of probable particle kinetic energies, and therefore must extend from $-\infty$ to $+\infty$, even though particles may not actually exist in this configuration. It is in this sense that the $\hat{v}_x$ axis differs from the actual particle spatial velocity $v_x$, even though they must be represented similarly. For convenience, we represent the $\tau v_x$ axis of the phase-space plane as $y$ and the spatial particle velocity as $v_x$ in order to indicate the discussed distinction between the two terms. In the succeeding analyses, the flow of phase-space will be analysed along the position and velocity axes of the phase-space.

\indent The Vlasov equation (\ref{Vlasov_equation}) can be represented in terms of the streaming function $\psi(x,v_x)$ using a stream matrix $\Psi$ as follows;
\begin{gather}\label{matrix_vlasov1}
\frac{\partial \eta}{\partial t} = \nabla \cdot\left( \Psi\nabla\eta\right),\\
\Psi(x,v_x) = \begin{pmatrix}
    0&\psi(x,v_x)\\
    -\psi(x,v_x)&0
\end{pmatrix} ,\\
\text{where,}\quad \nabla = \frac{\partial }{\partial x}\hat{x} + \frac{\partial}{\partial y}\hat{y}= \frac{\partial }{\partial x}\hat{x} + \frac{1}{\tau}\frac{\partial}{\partial v_x}\hat{v}_x.
\end{gather}
The form of Vlasov equation shown in equation (\ref{matrix_vlasov1}), describes, in fluid-analogous notation, a diffusion-free flow of the two-dimensional fluid \citep{Fannjiang1996DiffusionTurbulence}. 

Using the above relations, the Lobo-Sayal model presents a velocity field $\bm{\mathcal{V}}$ and a vorticity field $\bm{\xi}$ of the phase-space:
\begin{gather}\allowdisplaybreaks
    \bm{\mathcal{V}} = \frac{d}{dt}\left( x\hat x + \tau v_x \hat v_x\right) = v_x \hat{x} + \frac{\tau}{m}F(x) \hat{v}_x,\label{psveocityfield}\\
    \bm{\xi} = \nabla\times \bm{\mathcal{V}}=  \left(\frac{\tau}{m}\frac{\partial F}{\partial x}-\frac{1}{\tau}\right).\label{ps_vorticity_field}
\end{gather}
In the above equations, $F(x)$ represents the conservative force field experienced by the system. The probability distribution function $f(x,v_x)$ is treated as the particle density $\eta$ of the phase-space:
\begin{equation}\label{psdensity}
    \eta(x,v_x) = \frac{1}{\tau}f(x,v_x).
\end{equation}
Additionally, the phase-space hydrodynamic model also presents an interesting theory of a kinetic pressure formation in the phase-space of collision-less systems. This pressure, according to \citet{Lobo2024TheoryPlasmas}, develops due to an uneven probability distribution along position-space. Representing particle Hamiltonian as $H(x,v_x)$, the kinetic pressure $\bm{P}$ is defined as:
\begin{equation}\label{pressure_definition}
        \bm{P} = -\eta K_BT,
\end{equation}
assuming isothermal equilibrium of particles at some temperature $T$. The phase-space flow, due to its non-diverging nature, produces no net flux across a closed phase-space volume in the phase-space,
\begin{equation}\label{nodivergencecondition}
    \nabla\cdot\bm{\mathcal{V}}=0 \implies \iint\bm{\mathcal{V}}\cdot \tau dx dy = 0.
\end{equation}
This implies that there is no net flux of energy across the phase-space volume due to the flow of particles through it, which is the source of the phase-space pressure formation, as shown by \citet{Lobo2024TheoryPlasmas}. However, this pressure is observed upon restricting the analyses of the phase-space fluid flow along one axis. 

\subsection{Diffusion of the phase-space hydrodynamic fluid}

\indent In order to present the theory of probability density diffusion in phase-space, the Vlasov equation presented in equation (\ref{matrix_vlasov1}) can be modified and represented as:
\begin{gather}\label{modified_vlasov_eqn}
    \frac{\partial \eta}{\partial t} = \nabla \cdot\left[\left( \sigma + \Psi\right)\nabla\eta\right],\quad \text{where,}\\
\sigma(x,v_x) = \begin{pmatrix}
    \mathcal{F}(v_x)&0\\
    0&\mathcal{G}(x)
\end{pmatrix}.
\intertext{The respective coefficients $\mathcal{F}(v_x)$ and $\mathcal{G}(x)$ are defined as --}
\mathcal{F}(v_x) = \frac{1}{\eta}\frac{\partial^2\eta}{\partial y^2}=\frac{1}{\tau^2\eta}\frac{\partial^2\eta}{\partial v_x^2},\quad\mathcal{G}(x) = -\frac{1}{\eta}\frac{\partial^2\eta}{\partial x^2}.\label{diffusioncoeffs}
\end{gather}
It can be easily shown that the convection-diffusion equation (\ref{modified_vlasov_eqn}) readily expands into the ordinary Vlasov equation (\ref{Vlasov_equation}), and this has been shown in the Appendix section of this paper. The diffusion coefficients $\mathcal{F}(v_x)$ and $\mathcal{G}(x)$ shown above describe anisotropic diffusive behaviour of the phase-space evolution. Specifically for a Maxwellian distribution, these coefficients can be determined to be equal to --
\begin{gather}
\eta(x,v_x) =\frac{1}{N\tau}\exp\left[-\frac{H(x,v_x)}{K_BT}\right]\label{distributionmaxwellian}\\
    \mathcal{F}(v_x) = \frac{m/\tau^2}{( K_BT)^2}(mv_x^2 - K_BT)\label{fvformaxwellian}\\
    \mathcal{G}(x) = -\frac{q^2}{(K_BT)^2}\left( K_BT\frac{\rho(x)}{q\varepsilon_0} + E(x)^2 \right)\label{gxforaxwellian}.
\end{gather}
In the above equations, $N$ represents the norm of the normalised particle probability density and $\rho(x)$ represents the spatial charge density of the system.

 The above modified Vlasov equation (\ref{modified_vlasov_eqn}) represents phase-space dynamics in analogy to two-dimensional conventional fluids \citep{Fannjiang1996DiffusionTurbulence}. The diffusive flow of phase-space in such systems can then be segregated along each axis in order to display the convection-diffusion equation of phase-space along position and velocity axes. Setting the velocity axis $y$ to constant\footnote{This means we neglect the variation of particle density along the $y$ axis.}, we get --
\begin{equation}\label{convection-diffusionalongx}
    \frac{\partial \eta}{\partial t} = -v_x\frac{\partial \eta}{\partial x} + \mathcal{F}(v_x)\frac{\partial^2\eta}{\partial x^2}.
\end{equation}
Employing the Maxwellian distribution and averaging the above equation over the fluid element velocity $v_x$, we get --
\begin{equation}\label{convection-diffusionalongxaveraged}
    \frac{\partial \eta}{\partial t} = -\langle \nu_x\rangle\frac{\partial \eta}{\partial x} + \frac{m/\tau^2}{( K_BT)^2}(\langle mv_x^2\rangle - K_BT)\frac{\partial^2\eta}{\partial x^2}
\end{equation}
We now segregate the phase-space elements into two thermodynamically distinct species -- un-trapped phase-space fluid elements with positive Hamiltonian and the trapped elements with negative Hamiltonian $(-K_BT_t)$, where $T_t$ represents the temperature of the trapped particles in the system. In the un-trapped region, the diffusion coefficient inadvertently reduces to zero in equation (\ref{convection-diffusionalongxaveraged}). In the trapped region, however, the equation becomes --
\begin{equation}\label{betacontriinpositionspace}
     \frac{\partial \eta}{\partial t} = -\langle \nu_x\rangle\frac{\partial \eta}{\partial x} -\frac{m/\tau^2}{ K_BT}\left( 1+\frac{1}{\beta}\right)\frac{\partial^2\eta}{\partial x^2}
\end{equation}
Here, $\beta$ is the negative temperature ratio of the free particle temperature to the trapped particle temperature $(-T/T_t)$. This particle trapping parameter was first introduced by \citet{Schamel1979} in his study of electron phase-space hole distribution in phase-space for the trapped region and has been an ubiquitous parameter in the study of electron holes in collision-less plasmas, due to its significance in the nonlinear dispersion relation \citep{Schamel1979, schamel1986electron, schamel2020novel, Schamel2023PatternEquilibria, hutch2017}. Equation (\ref{betacontriinpositionspace}) therefore describes an interesting outcome -- diffusion in phase-space probability density occurs only when the particles shift from thermal equilibria, such as in regions of particle trapping. This is signified by the particle trapping $\beta$ parameter. In case of thermal equilibrium, there is no diffusion and hence, the contribution of the diffusion term in equation (\ref{modified_vlasov_eqn}) becomes zero.\\
\indent A similar conclusion can be inferred from the velocity-space diffusion coefficient $\mathcal{G}(x)$. In the presence of uniform particle distributions, both spatial charge density $\rho(x)$ and electric field $E(x)$ become zero and the diffusion vanishes. It occurs only when there is a reduction of a particle density in the phase-space region, thus causing the development of a local charge concentration (and associated phase-space vorticity concentration, as shown by equation \ref{ps_vorticity_field}) and an electrostatic field pressure.

\subsection{Turbulence in the phase-space hydrodynamic flow}
\indent Analogous to a two-dimensional fluid flow, the nature of the phase-space fluid flow can now be analyzed by comparing the inertial term $(v_x\partial_x\eta)$ and diffusion term $(\mathcal{F}(v_x)\partial^2_x\eta)$, in equation (\ref{convection-diffusionalongx}). A difference of the absolute form of the two terms, $\bm{\Theta}_x$ is presented which describes this comparison --
\begin{equation}
    \bm{\Theta}_x = |v_x\partial_x \eta|-|\mathcal{F}(v_x)\partial^2_x\eta|.
\end{equation}
The dominance of inertial flow over the diffusive flow can be determined from this difference in the condition that $\bm{\Theta} >0$, where as the $\bm{\Theta}< 0$ condition indicates presence of a strong diffusive (viscous) term. A similar analysis can then also be performed along the $y$ axis of the phase-space --
\begin{gather}
  \frac{\partial \eta}{\partial t}\Biggr|_{x=\text{constant}}= -\frac{q}{m}E(x)\nabla_y\eta +\mathcal{G}(x)\nabla_y^2\eta. \\
    \bm{\Theta}_y = |\frac{q}{m}E(x)\nabla_y\eta| -|\mathcal{G}(x)\nabla_y^2\eta|.
\end{gather}
  In analogy to conventional fluid flows, where inertial flow dominance over the viscous (diffusive) flow indicates turbulent flow, the condition of large, positive  $\bm{\Theta}_{x,y}$ indicates a turbulence in the phase-space fluid. This can be also confirmed by formation of phase-space vortices, similar to vortex formation in ordinary fluids due to turbulence. This parameter can be measured in numerical simulations of plasmas, and growth of turbulence and phase-space vortices can be observed in accordance with the above discussion. The mathematical analyses depicts a fluid-analogous diffusion of the phase-space probability density due existing to non-equilibrium conditions, and describe the role of the particle trapping parameter in determining this diffusive flow. Further, they present a technique for analytical determination of the nature of the phase-space fluid-like dynamics. This nature can, in fluid-analogous terms, be identified as turbulent-like or streamline-like forms.
  \begin{figure}[!ht]
    \centering
    \includegraphics[width=1\linewidth]{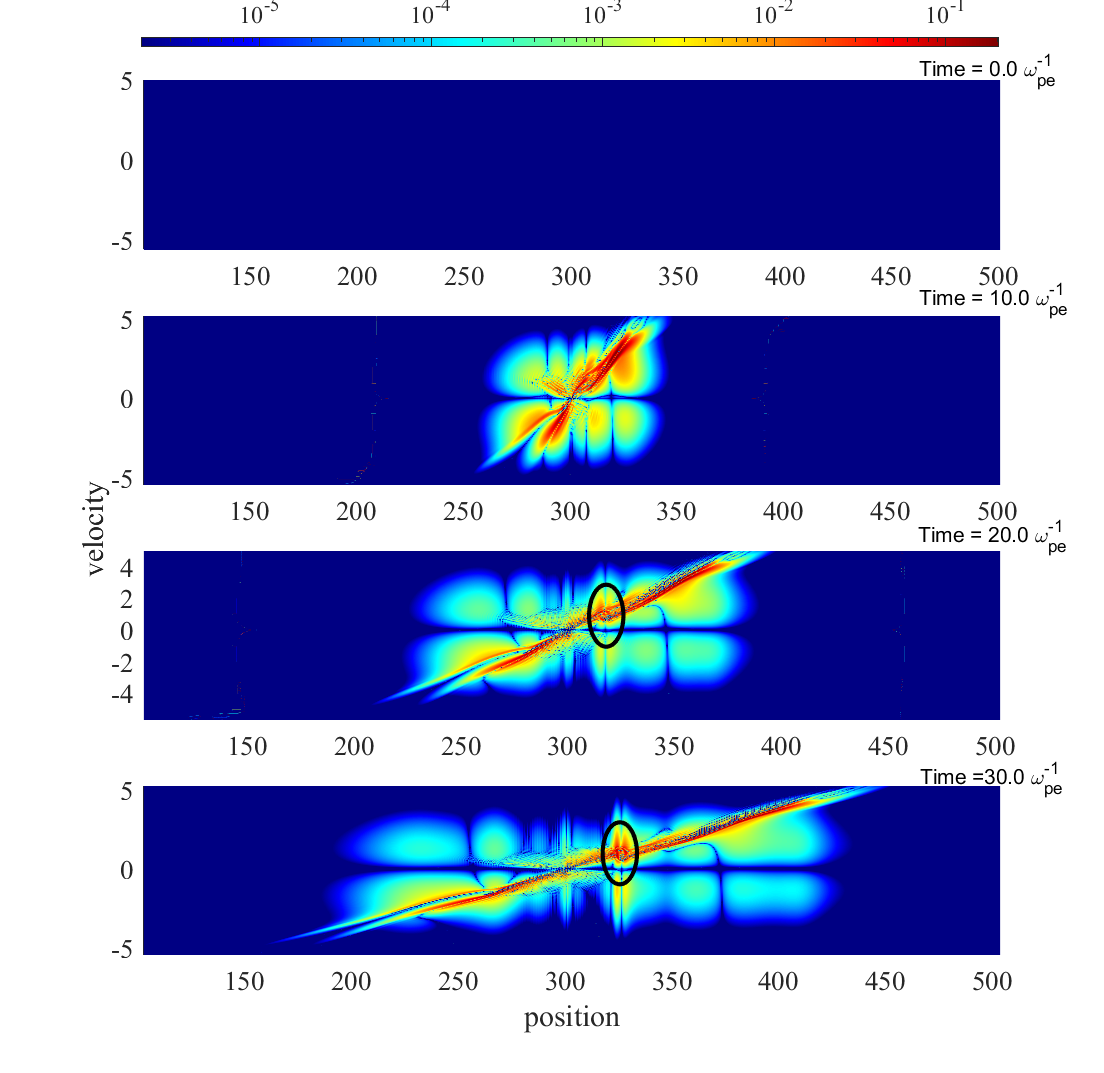}
    \caption{Formation of phase-space turbulence, depicted by the $\bm{\Theta}_x$ parameter (shown in logarithmic scale), resulting in the formation of a phase-space vortex (encircled in black). }
    \label{fig:turbulence_generation_elec_qm}
\end{figure}

  \section{Simulation Results and Discussions}
  
  \indent In reference to the mathematical analyses presented above, we present some numerical simulations of plasma systems which depict and confirm our findings. These simulation studies are performed using kinetic Vlasov technique and each employs the finite splitting scheme \citep{Cheng1976} for the numerical integration. First, we present the phase-space of a solitary electron hole in the cylindrically wave-guided experimental set-up \citep{Saeki1979, Lynov1979}. We consider a stationary and uniform ion background, along-with reflecting boundary conditions. A solitary electron hole is formed by introducing a step-like potential pulse \citep{turikov1978computer}. appropriately normalised units of position (with electron Debye length), time (with inversed electron plasma frequency), velocity (with thermal electron speed) and probability density are used.\\
  \begin{figure}[!ht]
  \centering
  \begin{subfigure}{1\linewidth}
      \includegraphics[width=1\linewidth]{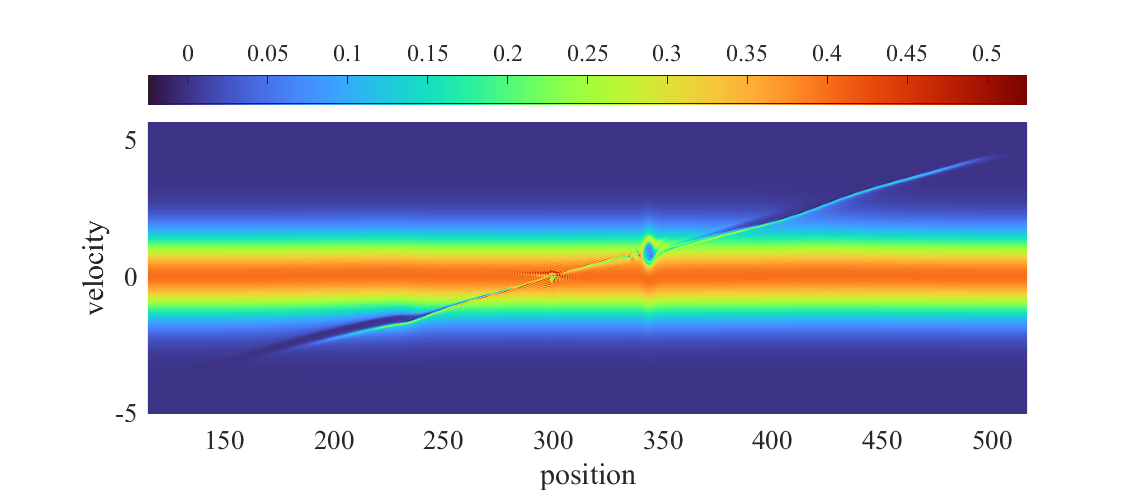}
      \caption{}
  \end{subfigure}
  \begin{subfigure}{1\linewidth}
      \includegraphics[width=1\linewidth]{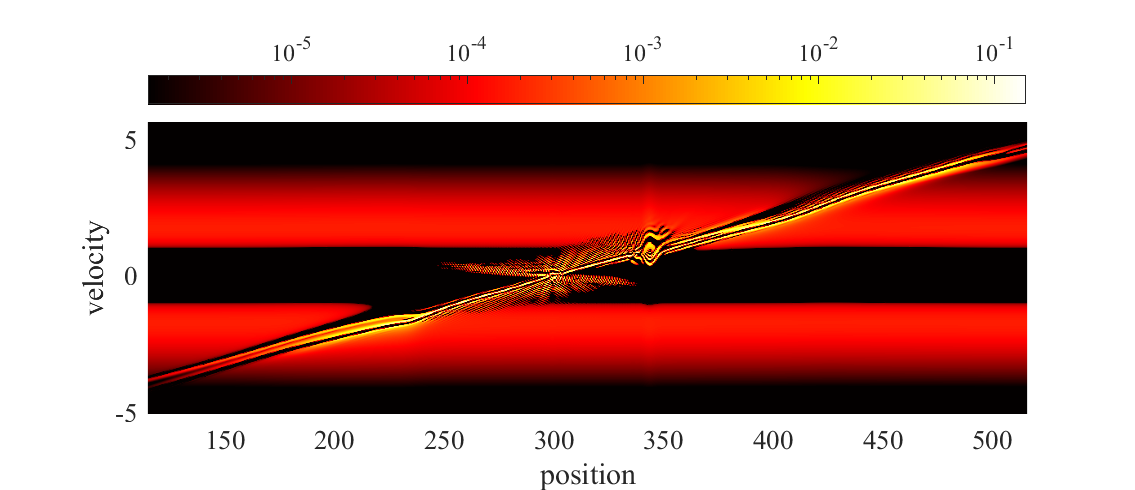}
      \caption{}
  \end{subfigure}
  \begin{subfigure}{1\linewidth}
      \includegraphics[width=1\linewidth]{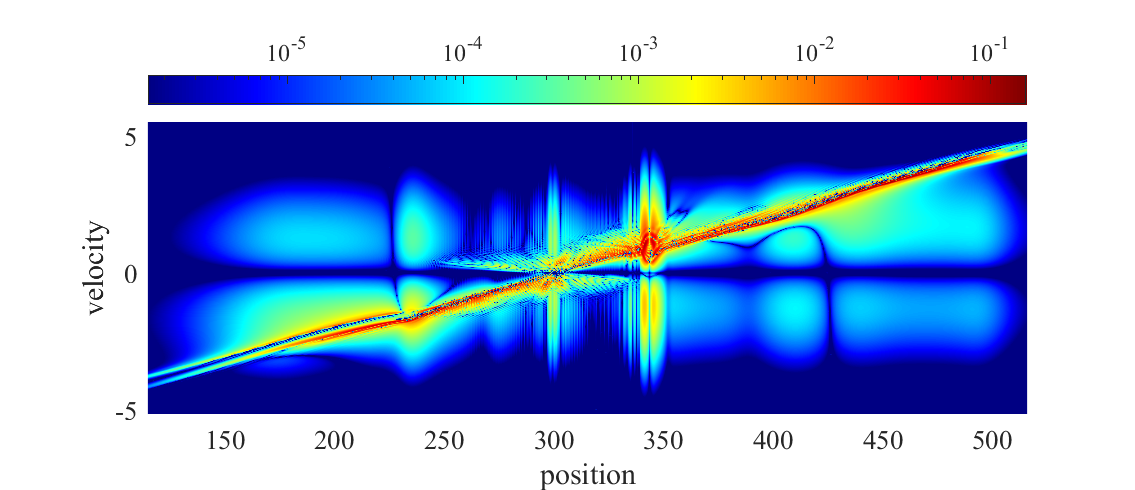}
      \caption{}
  \end{subfigure}
  \begin{subfigure}{1 \linewidth}
      \includegraphics[width=1\linewidth]{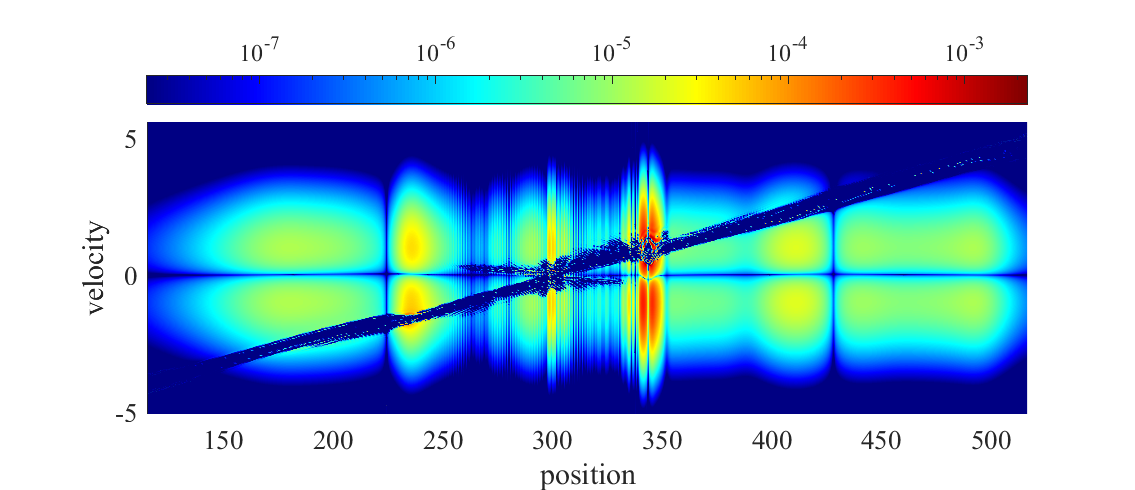}
      \caption{}
  \end{subfigure}
    \caption{An electron phase-space vortex formed in a cylindrically wave-guided set-up. (a) Electron phase-space density $f(x,v_x)$ depicting a hole. (b) The diffusion coefficient $\mathcal{F}_v\cdot\eta$ depicting presence of phase-space diffusion, and the positive values of (c) $\bm{\Theta}_x$ and the (d) $\bm{\Theta}_y$ parameters depicting conditions of turbulence with values greater than zero in the regions of the phase-space hole. Portraits (b), (c) and (d) in logarithmic scales.}
    \label{fig:qmachine_electron}
\end{figure}
\begin{figure}[!ht]
    \centering
    \begin{subfigure}{0.8\linewidth}
        \includegraphics[width=1\linewidth]{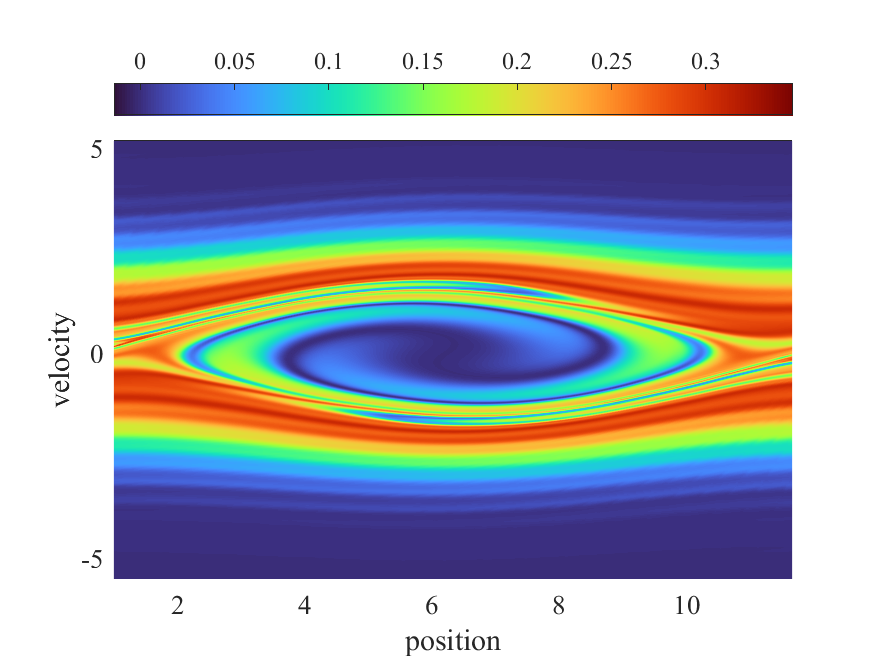}
        \caption{}
    \end{subfigure}
    \begin{subfigure}{0.8\linewidth}
        \includegraphics[width=1\linewidth]{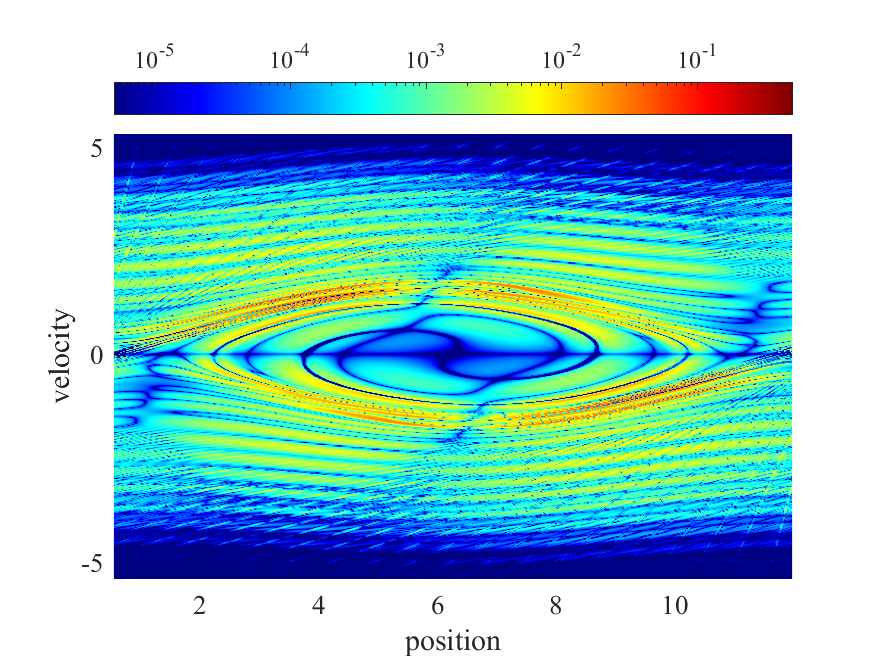}
        \caption{}
    \end{subfigure}
    \begin{subfigure}{0.8\linewidth}
        \includegraphics[width=1\linewidth]{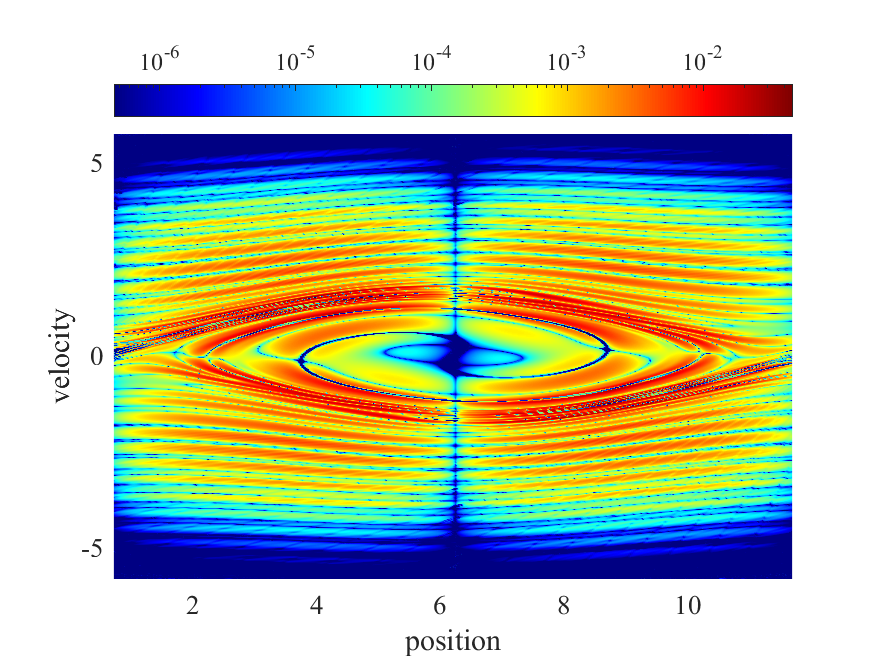}
        \caption{}
    \end{subfigure}
    \caption{Phase-space portraits of (a) probability distribution function, (b) $\bm{\Theta}_x$ and (c) $\bm{\Theta}_y$ in the region of a single electron hole formed from two-stream instability. Figures (a) in linear scale, (b) and (c) in logarithmic scale.}
    \label{fig:twostrelechole}
\end{figure}

\indent Figure \ref{fig:turbulence_generation_elec_qm} shows the development of phase-space turbulence due to the perturbation, leading to the development of the phase-space vortex. The growth and spread of the turbulence can be clearly seen, with concentration of the dominating inertial flow (over the phase-space diffusion) in the region of the phase-space vortex. Figure \ref{fig:qmachine_electron} depicts the presence of phase-space diffusion and turbulence at the cite of the electron hole. The presence of the spatial diffusion coefficient $\mathcal{F}_v$ is shown at the cite of the phase-space hole, which bears trapped electrons in the phase-space (see figure \ref{fig:qmachine_electron}b). The positive values of $\bm{\Theta}_{x,y}$ shown in figures \ref{fig:qmachine_electron}b and \ref{fig:qmachine_electron}c also depict inertial flow dominance in these regions, which in turn give rise to the phase-space vortex, in analogy to vortex formation in conventional fluids.\\
\begin{figure}[!hb]
    \centering
    \begin{subfigure}{1\linewidth}
    \includegraphics[width=1\linewidth]{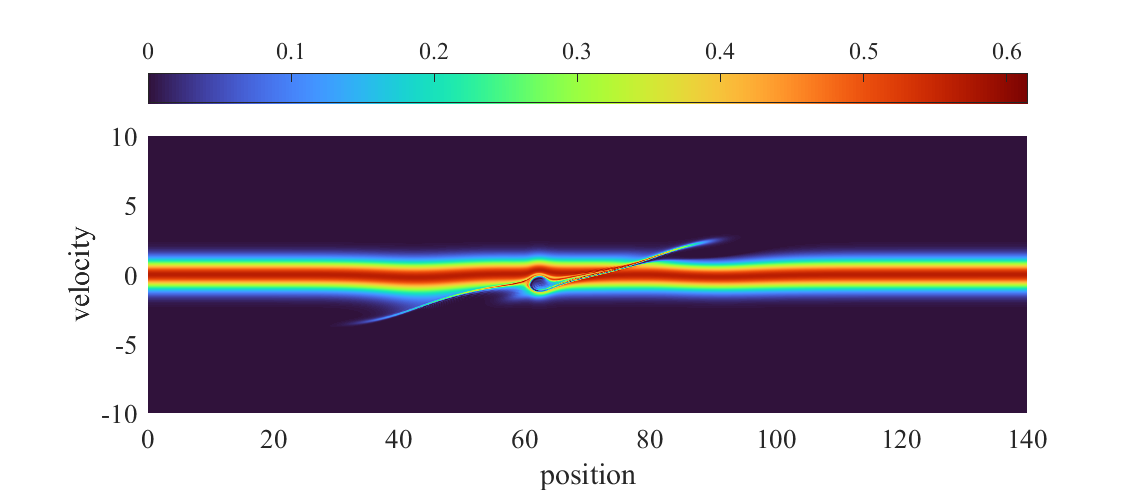}
    \caption{}    
    \end{subfigure}
    \begin{subfigure}{1\linewidth}
    \includegraphics[width=1\linewidth]{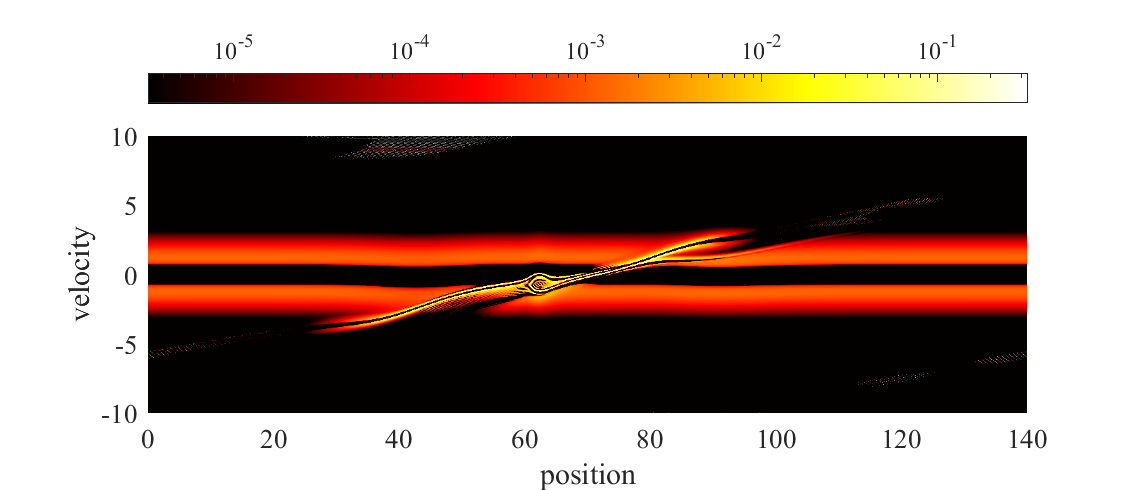}
    \caption{}    
    \end{subfigure}
    \begin{subfigure}{1\linewidth}
    \includegraphics[width=1\linewidth]{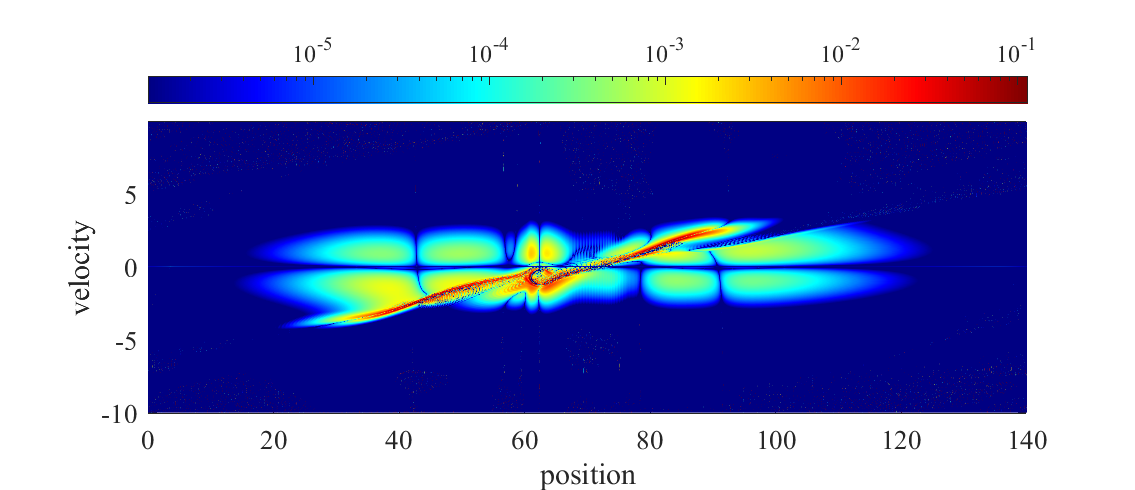}
    \caption{}    
    \end{subfigure}
    \caption{Ion phase-space portraits of (a) phase-space density $f(x,v_x)$, (b) spatial diffusion coefficient $\mathcal{F}(v_x)$ and (c) spatial turbulence parameter $\bm{\Theta}_x$. Figures (b) and (c) in logarithmic scales.}
    \label{fig:ionhole}
\end{figure}
\indent A similar analysis can be performed on an electron hole formed using two-stream instability, as shown in figure \ref{fig:twostrelechole}. For the same, periodic boundary conditions are employed in the numerical simulation of a two-stream plasma along-with an initial sinusoidal perturbation. The ion remains as a uniform stationary background and a two-beam Maxwellian electron distribution is used as the initial distribution. It can be seen that in the region of the electron hole, the phase-space flow is dominated by the inertial flows along both position and velocity axes. These phase-space holes, which have a characteristic property of being symmetric across the phase-space position $(x)$ axes, can be seen as regions of large $\bm{\Theta}_{x,y}$ as shown in figures \ref{fig:twostrelechole}b and \ref{fig:twostrelechole}c.\\
\indent In a similar approach, the formation of an ion phase-space vortex due to turbulence formation in the fluid ion phase-space can be demonstrated. For this, we use the simulation technique employed by \citet{Lobo2024AccelerationPlasma} for the formation of solitary ion holes. We assume thermal electron distribution $n_e = \exp(e\phi(x)/K_BT_e)$, ion-electron temperature ratio $T_i/T_e = 100.0$, and an initial Maxwellian distribution of ions. A perturbing potential step is introduced and the formation of an ion hole is observed. The ion phase-space is then analysed using techniques introduced in this paper. Figure \ref{fig:ionhole} showcases these observations. It can be seen that a solitary ion phase-space hole is formed, and associated to it is a region of concentrated phase-space diffusive flow. This flow is, however, dominated by the inertial flow of the phase-space, marked by the $\bm{\Theta}_x$ parameter. The presence of the phase-space vortex at this region of phase-space turbulence is clearly visible.\\
\indent The simulation results shown in figures \ref{fig:qmachine_electron}, \ref{fig:twostrelechole} and \ref{fig:ionhole} collectively confirm the mathematical analyses presented initially in this work. Specifically, they confirm that the phase-space of particles, in fluid-like analogy, present a diffusive flow along each axis of the phase-space, when considered independently. They show that the anisotropic diffusion coefficient can be determined for the phase-space regions and the inertial flow of the phase-space density can be compared with this diffusive flow. In some regions of the phase-space, the inertial flow of the fluid phase-space dominate this diffusion. This results in a turbulence-like state of the fluid phase-space analogue, which results in the formation of a phase-space vortex.

\section{Vorticity field transport in the phase-space}
With the development of the phase-space diffusion terms in $\sec$ \ref{sec2}, the phenomena of vorticity field $\bm{\xi}(x)$ transport in the hydrodynamic phase-space can be explored in terms of the well-known vorticity transport equation \citep{batchelor1967introduction}:
\begin{equation}\label{vorticity_transport_3d}
    \frac{\partial \bm{\xi}}{\partial t} + (\bm{\mathcal{V}}\cdot \nabla)\bm{\xi} = (\bm{\xi}.\nabla)\bm{\mathcal{V}} + \sigma\nabla^2\bm{\xi}.
\end{equation}
Equation (\ref{vorticity_transport_3d}) represents the generalised three-dimensional evolution equation of the vorticity field of a fluid or a fluid-analogous system, such as the hydrodynamic phase-space. It shows that the evolution of the fluid vorticity field is balanced by the vortex stretching term $((\bm{\xi}.\nabla)\bm{\mathcal{V}})$ against the diffusion term $(\sigma\nabla^2\bm{\xi})$. Restricting equation (\ref{vorticity_transport_3d}) for the two-dimensional phase-space surface, we get:
\begin{equation}\label{vorticity_transport_2d}
    \frac{\partial \bm{\xi}}{\partial t} + v_x\frac{\partial\bm{\xi}}{\partial x} = \sigma_{xx}\frac{\partial^2\bm{\xi}}{\partial x^2}.
\end{equation}
Here, the velocity-space derivatives are dropped since the vorticity field $\bm{\xi}(x)$ does not depend on the velocity coordinate $v_x$. Also, the $\sigma_{xx}$ term represents the averaged $\bm{\mathcal{F}}(v_x)$ term shown in equation (\ref{betacontriinpositionspace}), considering the effect of the trapped particles $(\beta \neq -1)$. The equation becomes:
\begin{gather}\label{vort}
    \frac{\partial \bm{\xi}}{\partial t} + v_x\frac{\partial\bm{\xi}}{\partial x} = \frac{m/\tau^2}{K_BT}\left(1+ \frac{1}{\beta}\right)\frac{\partial^2\bm{\xi}}{\partial x^2}.
    \intertext{Using equation (\ref{ps_vorticity_field}) to determine the vorticity field of the plasma species' phase-space and inserting into the vorticity transport equation,}
   \bm{\xi}(x)=\left(\frac{\tau}{m}\frac{\partial F}{\partial x}-\frac{1}{\tau}\right) = \left(\frac{q\tau}{m}\frac{\partial E}{\partial x}-\frac{1}{\tau}\right)=\left(\frac{q\tau}{m}\frac{\rho(x)}{\varepsilon_0}-\frac{1}{\tau}\right)
   \intertext{In order to further solve the problem, we consider the plasma to be composed on electrons and singly-charged, stationary ions distributed uniformly along space in order to provide a neutralising background, Therefore, $\tau$ becomes the inverse of electron plasma frequency,}
   \tau=\omega_{pe}^{-1} = \sqrt{\frac{\varepsilon_0 m_e}{n_0e^2}}.\\
   \therefore \quad\bm{\xi}(x) = \left(\frac{-e\tau}{m_e}\frac{\rho(x)}{\varepsilon_0} - \frac{1}{\tau}\right)=-\frac{1}{\tau}\left(\rho(x)+1\right).
   \intertext{Equation (\ref{vort}) therefore becomes,}
    \frac{\partial \rho}{\partial t} + v_x\frac{\partial\rho}{\partial x} = a\left(1+ \frac{1}{\beta}\right)\frac{\partial^2\rho(x)}{\partial x^2}.\label{vort_tran_step1}
\end{gather}
For simplicity, we represent $\frac{m_e/\tau^2}{K_BT_e}=a$. 

The equation (\ref{vort_tran_step1}) though seemingly similar to the linear Heat equation, is in-fact not so. It is a nonlinear second order differential equation where the nonlinearity results from the spatial velocity term $v_x$ which depends on the spatial potential field,
\begin{equation}
    v_x = \sqrt{\frac{2}{m}\left(\frac{m}{\tau}\psi-q\phi(x)\right)}.
\end{equation}
Here, $m\psi/\tau$ represents the particle Hamiltonian. Hence, it is possible during its mathematical treatment to map equation (\ref{vort_tran_step1}) to a dispersive form. Considering the no-divergence condition of the phase-space flow in equation (\ref{nodivergencecondition}).
\begin{gather}
    \nabla\cdot\bm{\mathcal{V}}=0\Rightarrow \frac{\partial v_x}{\partial x} + \frac{-e}{m_e}\frac{\partial E}{\partial v_x}\Rightarrow \frac{\partial v_x}{\partial x} = \frac{e}{m_e}\frac{\partial E(x)}{\partial v_x}.
    \intertext{Following chain-rule, we obtain the following:}
    \frac{\partial v_x}{\partial x} = \sqrt{\frac{e\rho(x)}{m_e\varepsilon_0}}.
    \intertext{The validity of this relation, as opposed to the intuitive idea that $v_x$ is independent of position space $(x)$ is shown in Appendix B. We now transform the co-ordinates of equation (\ref{vort_tran_step1}) in the frame of the travelling vorticity wave, with a phase-speed $M$. Therefore,}
    x\rightarrow x - Mt = \lambda,\quad
    \frac{\partial}{\partial x} \rightarrow \frac{d}{d\lambda},\quad
    \frac{\partial}{\partial t} = -M\frac{d}{d\lambda},\\
    -M\frac{d \rho}{d \lambda} + v_x(\lambda)\frac{d \rho}{d \lambda} = a\left(1+ \frac{1}{\beta}\right)\frac{d^2\rho}{d\lambda^2}.\label{subs}
    \intertext{ Differentiating equation (\ref{vort_tran_step1}) with respect to position $x$, we get:}
    -M\frac{d^2\rho}{d\lambda^2} + \frac{dv_x}{d\lambda}\frac{d\rho}{d\lambda} + v_x\frac{d^2\rho}{d\lambda^2} = a\left(1+\frac{1}{\beta}\right)\frac{d^3\rho}{d\lambda^3}.\label{subser}
    \intertext{Substituting the second order derivative from equation (\ref{subs}) into equation (\ref{subser}) gives:}
    \Rightarrow \frac{1}{\nu}(v_x-M)^2\left(\frac{d\rho}{d\lambda} \right) + \frac{d\rho}{d\lambda}\sqrt{\frac{e\rho(\lambda)}{m_e\varepsilon_0}} - \nu\frac{d^3\rho}{d\lambda^3}=0,
    \intertext{where,}
    \nu = a\left(\frac{1+\beta}{\beta}\right).
    \end{gather}
 Representing $v_x$ in terms of particle Hamiltonian $H$ and electrostatic potential $\phi(\lambda)$,
 \begin{multline}\label{vort_tr_2}
    \left(\sqrt{\frac{2}{m_e}(H+e\phi(\lambda))}-M\right)^2\frac{d\rho}{d\lambda} + \left(\nu\sqrt{\frac{e}{m_e\varepsilon_0}}\right) \sqrt{\rho(\lambda)}\frac{d\rho}{d\lambda} \\- \nu^2\frac{d^3\rho}{d\lambda^3}=0.
 \end{multline}
 Expanding equation (\ref{vort_tr_2}), we get --
\begin{multline}\label{vorticity_semi_final}
    \left(\frac{2}{m_e}(H+e\phi(\lambda)) + M^2 - 2M\sqrt{\frac{2}{m_e}(H+e\phi(\lambda))}\right)\frac{d\rho}{d\lambda} \\+ c \sqrt{\rho(\lambda)}\frac{d\rho}{d\lambda} - \nu^2\frac{d^3\rho}{d\lambda^3}=0,\quad c=\left(\nu\sqrt{\frac{e}{m_e\varepsilon_0}}\right).
\end{multline}

Equation (\ref{vorticity_semi_final}) is the vorticity transport equation, representing the form of the phase-space vorticity field formed due to particle trapping, and is dependent on the value of the particle Hamiltonian $H$. In order to accommodate the trapping condition, we restrict $H \in [-\phi(x),0[$, based on which further analyses of equation (\ref{vorticity_semi_final}) can be conducted.

Upon considering the minima of trapped species Hamiltonian $(=-e\phi(x))$ and inserting in equation (\ref{vorticity_semi_final}), the equation obtains the following form:
\begin{gather}\label{voreqn_h_minima}
     M^2\frac{d\rho}{d\lambda} + c \sqrt{\rho(\lambda)}\frac{d\rho}{d\lambda} - \nu^2\frac{d^3\rho}{d\lambda^3}=0.
     \intertext{Transforming back into stationary frame, we get:}
      \frac{\partial\rho}{\partial t} + \alpha \sqrt{\rho(x)}\frac{\partial\rho}{\partial x} + \gamma\frac{\partial^3\rho}{\partial x^3}=0, \label{S-kdv_form}
      \intertext{where,}
      \begin{split}
      \alpha = -\frac{m_e/\tau^2}{MK_BT_e}\cdot\left(\sqrt{\frac{e}{m_e\varepsilon_0}}\right)\cdot\left(\frac{1+\beta}{\beta}\right),\quad \text{and}\\
      \gamma = \frac{1}{M}\left[\left(\frac{m_e/\tau^2}{K_BT_e}\right)\cdot\left(\frac{1+\beta}{\beta}\right)\right]^2.
      \end{split}
\end{gather}

Equation (\ref{S-kdv_form}) represents a well-known form of solitary-wave propagation, known as the Schamel-Korteweg-De Vries (S-KdV) equation, introduced initially by \citet{Schamel1972, Schamel1973AElectrons}. This form has been analytically explored by many authors \citep{Chatterjee2023ExplicitTransformation, Tariq2022ExplicitCoefficients, Khater2022Two-componentSpeed, Alquran2024DerivationEquations, Flamarion2024Non-integrableFramework, Yepez-Martinez2022NewEquation, Tariq2022OnApproaches, Soliman2023EffectPlasmas, AlAtawi2017ExactEquation, Lee2011ExactEquation, Yuan2013AllModels, Giresunlu2017OnEquation, Mushtaq2012DustElectrons}. 

Upon employing the $H\rightarrow 0^-$ case, equation (\ref{vorticity_semi_final}) retains a generalised S-KdV form, with some coefficients of spatial derivatives depending on the electrostatic potential $\phi(x)$ associated with the vorticity field:
\begin{multline}
    -\frac{2e}{Mm_e}\phi(\lambda)\frac{d\rho}{d\lambda} - M\frac{d\rho}{d\lambda} + 2\sqrt{\frac{2}{m_e}e\phi(\lambda)}\frac{d\rho}{d\lambda} \\+ \alpha \sqrt{\rho(\lambda)}\frac{d\rho}{d\lambda} + \gamma\frac{d^3\rho}{d\lambda^3}=0.
\end{multline}
Changing into stationary frame coordinates, we get:
\begin{multline}\label{skdvwithphi}
    \frac{\partial \rho}{\partial t} + \left(2\sqrt{\frac{2}{m_e}e\phi(x)}-\frac{2e}{Mm_e}\phi(x)\right)
\frac{\partial \rho}{\partial x} \\+ \alpha \sqrt{\rho(x)}\frac{\partial \rho}{\partial x} + \gamma\frac{\partial^3 \rho}{\partial x^3}=0.
\end{multline}
Equation (\ref{skdvwithphi}) represents the generalised S-KdV form of the Vorticity transport equation with coefficients which depend on the electrostatic potential $(\phi(x))$.

In this work, we omit exploring the solutions of equations (\ref{vorticity_semi_final}, \ref{S-kdv_form} \& \ref{skdvwithphi}), highlighting the conclusion that the vorticity transport equation, in its S-KdV form, presents vorticity wave-modes which must exhibit solitary-like behaviours, as is expected from the general solutions of the S-KdV equation. We leave the objective of determining these exact solutions of the vorticity transport equations (\ref{vorticity_semi_final}, \ref{S-kdv_form} \& \ref{skdvwithphi}) for our future endeavours in this field and proceed to the concluding remarks of our works in this paper.

\section{ Conclusion}
In this work, we have attempted to present an interesting outcome of the fluid-phase-space model -- the phase-space, analogous to conventional fluids, exhibits fluid-like diffusive flow. This flow can be described using anisotropic diffusion coefficients, which can be integrated into the classical Vlasov equation. The modified Vlasov equation then can be used to assess the convective-diffusive flow along each axis. Upon doing so, the phase-space reveals regions where the inertial flux of the phase-space fluid dominates significantly over the diffusive flow, causing a turbulent-like state and producing phase-space vortices. The presence of phase-space density diffusion also determines the nature of vorticity field transport in the phase-space. The vorticity transport equation modifies into the S-KdV form, which is well-known to produce solitary travelling wave solutions. \\
\indent The analysis presented in this work describes the phase-space vortices (electron and ion holes) as consequences of the nature of the flow of the phase-space fluid. In fluid-analogous study of the phase-space, it has been shown that the vortices in phase-space are similar in their evolution to that in conventional fluids, arising from regions of reduced diffusive flow. Phase-space holes -- nonlinear kinetic species born out of various trapping scenarios resulting from wave-particle interactions in collision-free plasmas, can also be addressed using the fluid-analogous approach to the phase-space. The work presents an answer to the problem initially presented in the introduction to this work -- phase-space vortices emerge out of turbulence-like states of the fluid-phase-space. It also identifies the role of the particle trapping parameter in the formation of this turbulent state. From equation (\ref{betacontriinpositionspace}) it can be seen that turbulence occurs along the position axis when the system shifts locally into thermodynamical non-equilibrium, which is represented by the $\beta$ parameter. This shift is responsible for both the diffusive nature of the probability density as-well-as its turbulent-like nature, which results into hole formation. A hole, which in terms of probability density is a state with comparatively reduced probability than its neighborhood in phase-space, emerges out of this turbulent state. By turbulence, it is meant that the inertial flow of the phase-space probability density, which is represented by the classical Vlasov term, dominates the diffusion term along the respective axis. This can be easily shown to be related to larger values of the $\beta$ parameter, therefore signifying a greater shift from thermodynamic equilibrium.\\
\indent Through this work, an attempt has been made to showcase the fluid-analogous nature of probability phase-space of many-particle systems, and its efficacy towards the study of various aspects of phase-space dynamics. The simulations of phase-space holes formed using various plasma processes presented in this work, and their agreement to the analyses presented herein, confirm the analytical findings of the fluid-analogous approach to the study of these kinetic structures. To conclude, we state that the fluid-analogous nature of the phase-space probability density and their relation to the spatial phenomena occurring in such systems, such as the problem of electron and ion holes presented in this work, present an interesting class of mathematical analyses and provide interesting outcomes which can be explored using the mathematical techniques of the fluid phase-space model \citet{Lobo2024TheoryPlasmas}. We rest our investigation on this note.

\section*{Code and Data Availability Statement}
The MATLAB code and the numerical data used for generating plots used in this work and presented analyses are available with the corresponding author (A. Lobo) upon reasonable request.
\section*{Appendix}
\subsection*{A: Expansion of the convection-diffusion equation and derivation of the Vlasov equation}
\renewcommand{\theequation}{A.\arabic{equation}}
\setcounter{equation}{0}
Upon expanding the phase-space diffusion-advection equation (\ref{modified_vlasov_eqn}), we get 
\begin{gather}
\frac{\partial \eta}{\partial t} = \begin{pmatrix}
    \partial_x & \partial_y
\end{pmatrix}\cdot \begin{pmatrix}
    \mathcal{F}(v_x)\partial_x\eta + \psi\partial_y\eta\\
    -\psi\partial_x\eta + \mathcal{G}(x)\partial_y\eta
\end{pmatrix}\\
 = \frac{\partial}{\partial x}\left(  \mathcal{F}(v_x)\partial_x\eta + \psi\partial_y\eta
 \right) + \frac{\partial}{\partial y}\left( -\psi\partial_x\eta + \mathcal{G}(x)\partial_y\eta \right).
 \end{gather}
 This expands into:
 \begin{multline}
    \frac{\partial \eta}{\partial t}   = \mathcal{F}(v_x)\frac{\partial^2\eta}{\partial x^2} + \frac{\partial \psi }{\partial x}\frac{\partial \eta}{\partial y}  + \psi\frac{\partial^2\eta}{\partial x\partial y} \\- \frac{\partial \psi}{\partial y}\frac{\partial \eta}{\partial x} -\psi\frac{\partial^2\eta}{\partial x\partial y} + \mathcal{G}(x)\frac{\partial^2\eta}{\partial y^2},
 \end{multline}
which readily reduces to the Vlasov equation --
\begin{equation}
    \frac{\partial \eta}{\partial t} =  \frac{\partial \psi }{\partial x}\frac{\partial \eta}{\partial y}  - \frac{\partial \psi}{\partial y}\frac{\partial \eta}{\partial x}.
\end{equation}
\subsection*{B: Validity of the $\partial_x v_x\neq 0 $ condition}
\renewcommand{\theequation}{B.\arabic{equation}}
\setcounter{equation}{0}
As discussed in $\S$ \ref{sec2}, the velocity field $\hat{x}\cdot\bm{\mathcal{V}}$ of the phase-space fluid is independent of the velocity axis $(\hat v _x = y)$, which extends between $\pm\infty$. The kinetic fluid density $\eta$ in the phase-space plane varies along the co-ordinates, and flows in the phase-space $xy$ plane with velocity field $\bm{\mathcal{V}}$. The velocity axis $y$ merely describes the extents of the probable particle velocities in the system. But the actual particle velocities, which in the phase-space is represented by the fluid velocity field $\bm{\mathcal{V}}\cdot\hat x$, is decided by the particle Hamiltonian:
\begin{gather}
    \bm{\mathcal{V}}\cdot \hat x = v_x = \sqrt{\frac{2}{m}(H-q\phi(x))}.\\
    \therefore\quad \frac{\partial v_x}{\partial x} = -q\sqrt{\frac{1}{2m}\left(\frac{1}{H-q\phi(x)}\right)}\frac{\partial \phi}{\partial x}.
\end{gather}
Therefore, the spatial dependence of the kinetic fluid exists in regions where the electrostatic field exists, and is zero only when $\partial_x\phi(x)=0$.

\bibliographystyle{apalike} 
\bibliography{references}

 \subsection*{Acknowledgement}
 One of the authors (A.L.) acknowledges the TMA-PAI PhD research fellowship provided by Sikkim Manipal University.

 \textbf{Competing interests:} The authors declare no competing interests. 
\indent \noindent \textbf{Correspondence} and request for materials should be addressed to A.L.  

\end{document}